\documentclass[11pt,twoside]{article}
\usepackage{asp2010}

\usepackage{graphicx}
\usepackage[usenames]{color}	
\usepackage{epstopdf}	

\newcommand{\numax}{\nu_{\mathrm{max}}}
\newcommand{\nuc}{\nu_{\mathrm{c}}}

\newcommand{\deriv} [2] {\frac {\textrm{d} #1 } {\textrm{d} #2} }

\newcommand{\eq}[1] {Eq.\,(\ref{#1})}
\newcommand{\eqn} [1] {
\begin{equation} #1
\end{equation}}

\newcommand{\dnuas}{\Delta \nu_{\mathrm{as}}}
\newcommand{\dnuobs}{\Delta \nu_{\mathrm{obs}}}

\def\apjl{ApJ}%
\def\apjs{ApJS}%
\def\aap{A\&A}%
\resetcounters

\begin{document}

\title{On the seismic scaling relations $\Delta \nu - \bar{\rho}$ and $\numax-\nuc$}
\author{K. Belkacem$^1$, R. Samadi$^1$, B. Mosser$^1$, M.J. Goupil$^1$ and H.-G. Ludwig$^{2,3}$
\affil{$^1$LESIA, UMR8109, Observatoire de Paris, Universit\'e Pierre et Marie Curie, 
Universit\'e Denis Diderot, 92195 Meudon Cedex, France.}
\affil{$^2$Zentrum f\"ur Astronomie der Universit\"at Heidelberg, Landessternwarte, K\"onigstuhl 12, D-69117 Heidelberg, Germany}
\affil{$^3$GEPI, Observatoire de Paris, CNRS UMR 8111, Universit\'e Denis Diderot, 5 Place Jules Janssen, 92195 Meudon Cedex, France}}

\begin{abstract}
Scaling relations between asteroseismic quantities and stellar parameters are essential tools for studying  stellar structure and evolution.  
We will address two of them, namely, the relation between the large frequency separation ($\Delta \nu$) and the mean density ($\bar{\rho}$) as well as the relation between the frequency of the maximum in the power
spectrum of solar-like oscillations ($\numax$) and the cut-off frequency ($\nuc$). 

For the first relation, we will consider the possible sources of uncertainties and explore them with the help of a grid of stellar models. For the second one, we will show that the basic physical picture is understood and that departure from the observed relation arises from the complexity of non-adiabatic processes involving time-dependent treatment of convection. This will be further discussed on the basis of a set of 3D hydrodynamical simulation of surface convection.
\end{abstract}

\section{Introduction}
\label{intro}

The advent of space-borne asteroseismology has been possible with the space missions CoRoT \citep{Baglin2006a,Baglin2006b,Michel2008} and {\it Kepler} \citep{Borucki2010}. These two missions have provided a wealth of high-quality and long-duration observational data that enable us to measure and characterize stellar oscillations. Among others, stars pulsating with solar-like oscillations (\emph{i.e.} excited and damped by the upper-most convective layers of low-mass stars) are particularly important since they show a rich spectrum allowing for a probe of their internal structure \citep[e.g.][]{Goupil2011a,Goupil2011b,JCD2012}. 

\begin{figure}
\centerline{\includegraphics[height=6.5cm,width=10cm]{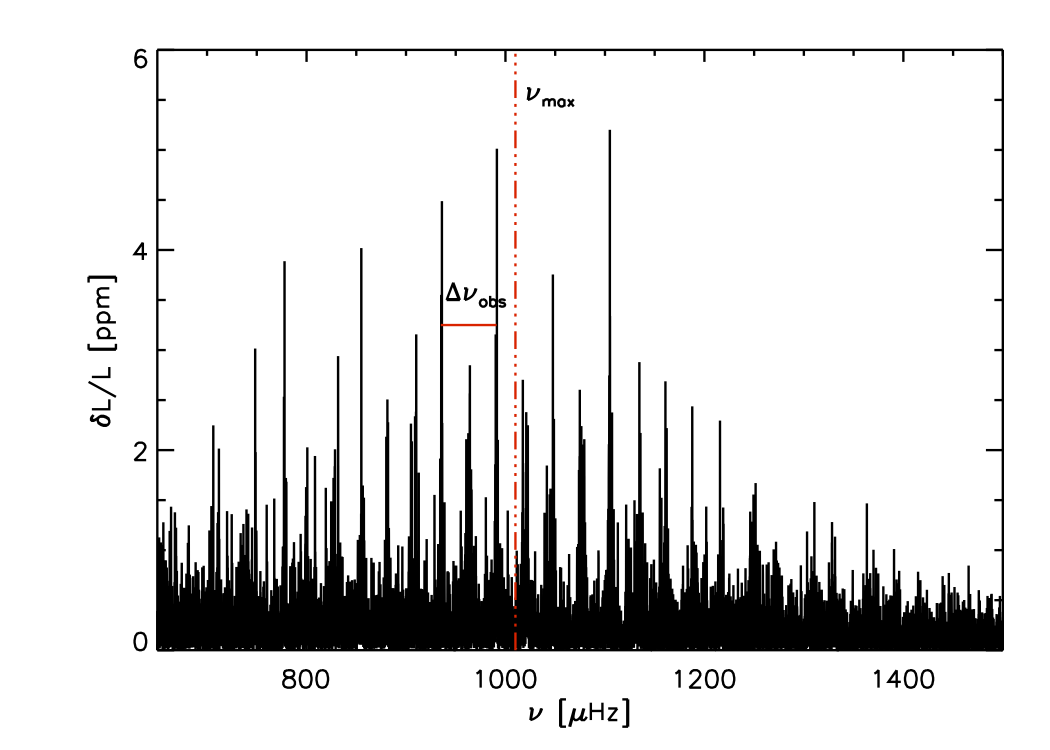}}
\caption{Power spectrum as a function of frequency of the star HD49385 observed by CoRoT during a period of 137 days \citep[see][for details]{Deheuvels2011}. The vertical red dotted-dashed line represents the position of $\nu_{\rm max}$ while the red solid horizontal segment illustrates the large separation ($\Delta \nu_{\rm obs}$) between two modes of same angular degree. }
\label{spectre}
\end{figure}

Up to now, several hundreds of main-sequence stars with solar-like oscillations have been detected, as well as several thousands oscillating red giant stars. These observation permit statistical analysis and gave birth to the \emph{ensemble asteroseismology}. This new approach is allowed by the large-scale exploitation of \emph{seismic indices} that are also called global seismic parameters. As depicted in Fig.~\ref{spectre}, one can easily identify two major seismic indices, namely: 
\begin{itemize}
\item the large separation, $\Delta \nu  \equiv \langle \nu_{n+1,\ell} - \nu_{n,\ell}\rangle$ (where $\nu_{n+1,\ell}$ is the frequency, $n$ is the radial order, $\ell$ is the angular degree, and $\langle \cdot \rangle$ stands for an average over frequency.). 
\item the frequency of the maximum height in the power spectrum, $\nu_{\rm max}$, which is defined such as $H \left(\nu_{\rm max} \right) = max \left[H \left(\nu \right) \right]$, where $H$ is the height in the power spectrum corrected from the background\footnote{We note that alternative definitions of $\nu_{\rm max}$ are possible and in particular in terms of mode amplitude rather than mode height \cite[see][for a discussion]{Belkacem2012b}, but for consistency with the theoretical results of \cite{Belkacem2011} we adopt this definition.}.
\end{itemize}
Although other scaling relations are available, we will restrict our focus to these quantities in this article. 
 
The determination of seismic indices provides a wealth of information since they can be related to global stellar parameters through \emph{scaling relations}. One can define them as relations between seismic indices and global stellar parameters such as the mass, radius, or effective temperature. The two commonly used relations are \citep[e.g.,][]{Ulrich86,Brown91,KB95,Belkacem2011}:  
\begin{itemize}
\item the relation between $\Delta \nu$ and the squared mean density of the star, \emph{i.e.}
\begin{equation}
\label{scaling_deltanu}
\Delta \nu  \propto \bar{\rho}^{1/2} \equiv \left(\frac{3}{4 \pi} \frac{M}{R^3}\right)^{1/2} \, ,
\end{equation}
where $\bar{\rho}$ is the mean density, $M$ is the total mass of the star, and $R$ is total radius. 
\item The second one relates the frequency of the maximum height in the power spectrum to the cut-off frequency,  \emph{i.e.}
\begin{equation}
\label{scaling_numax}
\nu_{\mathrm{max}} \propto \nu_{\mathrm{c}} \propto \frac{g}{\sqrt{T_{\rm eff}}} \propto \frac{M}{R^2 \sqrt{T_{\rm eff}}} \, , 
\end{equation}
where $\nu_{\rm c}$ refers to the cut-off frequency, \emph{i.e.} the frequency above which there is no more total reflection at the star surface. 
\end{itemize}
From \eq{scaling_deltanu} and \eq{scaling_numax}, the potential of ensemble asteroseismology immediately arises because one can  derive an estimate of stellar masses and radii (or alternatively mean densities and surface gravities) provided a determination of the effective temperature is available
\begin{eqnarray}
\label{relations_canoniques1}
\frac{M}{M_\odot} &\propto&  \left(\frac{\numax}{\numax^\odot}\right)^{3} \, \left(\frac{\Delta \nu}{\Delta \nu_\odot}\right)^{-4} \, \left(\frac{T_{\rm eff}}{T_{\rm eff,\odot}}\right)^{3/2} \, , \\
\label{relations_canoniques2}
\frac{R}{R_\odot} &\propto& \left(\frac{\numax}{\numax^\odot}\right) \, \left(\frac{\Delta \nu}{\Delta \nu_\odot}\right)^{-2} \, \left(\frac{T_{\rm eff}}{T_{\rm eff,\odot}}\right)^{1/2} \, ,
\end{eqnarray}
where $T_{\rm eff}$ is the effective temperature, and the symbol $\odot$ denotes the solar reference values. Equations (\ref{relations_canoniques1}) and (\ref{relations_canoniques2}) are the cornerstones of ensemble asteroseismology and provide a wealth of constraints on stellar structure and evolution, as well as stellar populations \cite[see][for a recent review]{Chaplin2013}. 

In this article, our objective is to investigate the physical foundations of this scaling relations. Indeed, many efforts are currently being undertaken to calibrate and validate these relations \citep[e.g.][]{Bedding2011,Bruntt2011,Miglio12b,Huber2012,SilvaAguirre2012} but a firm theoretical ground is certainly the best way to ensure their proper and well-motivated use. Such theoretical investigations have been previously performed  \citep[e.g.][]{Stello2009,White2011,Belkacem2011,Montalban2012,Miglio2012,Montalban2013}, but there is still a large gap to  fill to reach a full and satisfying understanding. We will therefore discuss the fundamental physical concept underlying the scaling relations presented in Eqs.~(\ref{relations_canoniques1}) and (\ref{relations_canoniques2}). In Sect.~\ref{sect:deltanu}, we address   the $\Delta \nu-\bar{\rho}$ relations and, with the help of stellar modeling, we discuss the accuracy of this relation. Section~\ref{sect:numax} will be dedicated to the $\numax-\nu_{\rm c}$ scaling relation by discussing its  theoretical background with the help of a set of 3D hydrodynamical simulations. Finally, concluding remarks are provided in Sect.~\ref{sect:conclu}. 

\section{The $\Delta \nu-\bar{\rho}$ scaling relation}
\label{sect:deltanu}

\begin{figure}
\centerline{\includegraphics[height=7.5cm,width=11cm]{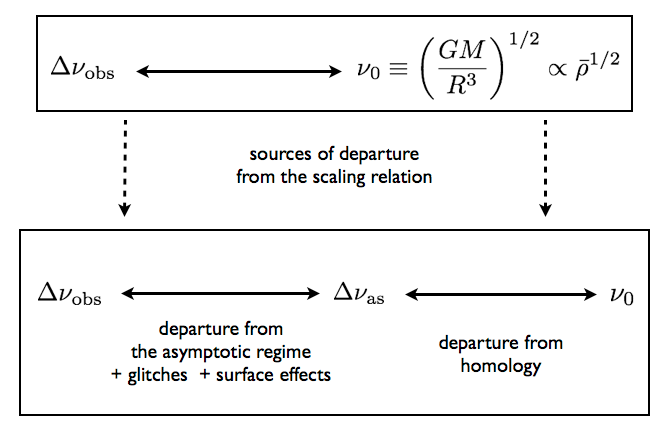}}
\caption{Sketch that illustrates the relation between the large separation and the mean density (top panel). The possible sources of departure from the scaling relation are displayed in the bottom panel, with particular emphasis on the  intermediate quantity $\Delta \nu_{\rm as}$ (defined in Sect.~\ref{definitions_prealables}). }
\label{schema_deltanu}
\end{figure}

In this section, we discuss the possible sources of departure from the $\Delta \nu-\bar{\rho}$ relation. But before going through the details, it is worth to make clear some definitions, summarized in Fig.~\ref{schema_deltanu}. 

\subsection{Preliminary definitions}
\label{definitions_prealables}

From an observational point of view, the \emph{observed} large separation is the near regularity measured around $\nu=\nu_{\rm max}$ because it is obviously the frequency for which the signal-to-noise ratio is maximum. It corresponds to radial order of about 20 for main-sequence stars down to two near the tip of the red giant branch. Indeed, as a low-mass star evolves, it exhibits excited solar-like oscillations at lower radial orders. We therefore define the large separation measured near $\nu_{\rm max}$ as $\Delta \nu_{\rm obs}$ \cite[see][for a more detailed discussion on $\Delta \nu_{\rm obs}$]{Mosser2013}. Note that, in this article,  $\Delta \nu_{\rm obs}$ is obtained by using computed frequencies as a proxy of measured frequencies. 

On the other hand, from a theoretical point of view, the large separation is a quantity introduced from an asymptotic analysis \citep[e.g.][]{Tassoul80,Gough1990} and as such it is formally valid at high radial orders. In other words, the requirements is that the characteristic vertical wavelength of the eigenfunctions must be small compared to the scale of variation of the equilibrium state (\emph{i.e.} $k_r \, H_{\{p,\rho,T\}} \gg  1$, where $k_r$ is the radial wavenumber of the oscillation, $H_{\{p,\rho,T\}}$ are the pressure, density, and temperature scale heights, respectively). 
Under this assumption, the large separation becomes   
\begin{equation}
\label{delta_nu_asymp}
\Delta \nu_{\rm as} \equiv \left( 2 \int_{0}^{R} \frac{{\rm d}r}{c_s}  \right)^{-1} = (2 \tau)^{-1} \, , 
\end{equation}
where $c_s$ is the adiabatic sound speed, and the quantity $\tau$ is the acoustic radius. 

Except the aforementioned large separations, one can also define $\nu_0$ such as \begin{equation}
\nu_0 \equiv \left( \frac{GM}{R^3} \right)^{1/2} \propto \bar{\rho}^{1/2} \, .
\end{equation}
This quantity is the inverse of the dynamical time-scale of a star, and as such can be called the dynamical frequency. It is then related to the mean density ($\bar{\rho}$). 

\subsection{Theoretical background of the $\Delta \nu_{\rm as} - \nu_0$ relation: homology}
\label{homology}

As displayed in Fig.~\ref{schema_deltanu}, the relation between $\Delta \nu_{\rm obs}$ and $\nu_0$ can be decomposed to introduce, as an intermediate, the quantity $\Delta \nu_{\rm as}$. Such a decomposition is useful since it makes clear the possible sources of departure from a perfect scaling (see Sect.~\ref{departure_delta}). It leads to identify that the main physical assumption is the homology between $\Delta \nu_{\rm as}$ and $\nu_0$.  

Let us consider two homologous stars, such that for two shells verifying $r/R = r^\prime/R^\prime$, the corresponding mass shells are equal ($m/M = m^\prime/M^\prime$), where $M, M^\prime$ are the total masses of two stars belonging to a homologous series, and $R,R^\prime$ their total radii. Under this assumption, it is possible to show from the equations of mass continuity and conservation of momentum that the sound speeds of both models are related by \citep[see for instance Sect. 20.1 of][]{Kippenhahn90}
\begin{equation}
\label{sound_homologous}
\frac{c_s}{c_s^\prime} =  \left( \frac{M}{M^\prime} \right)^{1/2} \,  \left( \frac{R}{R^\prime} \right)^{-1/2} \, . 
\end{equation}
Using Eq.~(\ref{sound_homologous}) together with the relation $r/R = r^\prime/R^\prime$, it is straightforward to demonstrate the desired scaling relation, \emph{i.e.}
\begin{equation}
\label{ratio}
\frac{\Delta \nu_{\rm as}}{\Delta \nu_{\rm as}^\prime}  = \left[\int_{0}^{R^\prime} \frac{\mathrm{d}r^\prime}{c_{s}^\prime}\right] \left[ \int_{0}^{R} \frac{\mathrm{d}r}{c_{s}}\right]^{-1} 
=  
\left(\frac{R^\prime}{R}\right)^{3/2}  \left(\frac{M}{M^\prime}\right)^{1/2} \, .
\end{equation}
In other words, if one of the considered star is the Sun, one has 
\begin{equation}
\label{finale_scale1}
\Delta \nu_{\rm as} = \left(\frac{\bar{\rho}}{\bar{\rho}_\odot}\right)^{1/2} \, \Delta \nu_{\odot, {\rm as}} \; , 
\end{equation}
 
Equation (\ref{finale_scale1}) demonstrates the scaling relation between $\Delta \nu_{\rm as}$ and $\bar{\rho}$, and shows that the underlying hypothesis is homology. This assumption is in general considered as a crude one, which is however useful to get some insight into more complex models obtained from a full numerical computation.  As discussed extensively for instance by \cite{Kippenhahn90}, the main physical requirements for complete homology to holds are 
\begin{itemize}
\item Complete equilibrium, \emph{i.e.} that both stars must be in both hydrostatic and thermal equilibrium. 
\item The mean Rosseland opacities, the equation of state, and the energy generation rate must be power laws of density and temperature (or equivalently pressure and temperature). 
\item The physical mechanism that transports energy (convection or radiation) must be the same in the  considered stars.  
\end{itemize}
When applying the $\Delta \nu_{\rm as} - \nu_0$ relation between two low-mass stars from the main-sequence to the red giant branch as well as the red clump, all those requirements are violated. 
Nevertheless, one must keep in mind that mainly the upper layers will contribute to $\Delta \nu_{\rm as}$ since the inverse of the sound speed is higher in those layers (see Eq.~\ref{delta_nu_asymp}). Therefore, the consequences of the violation of the aforementioned requirements are not obvious, and as we will show and quantify in Sect.~\ref{departure_delta}, the $\Delta \nu_{\rm as} - \nu_0$ relation is quite accurate.  

\subsection{Sources of departure from the scaling relation}
\label{departure_delta}

\begin{figure}
\centerline{\includegraphics[width=10cm]{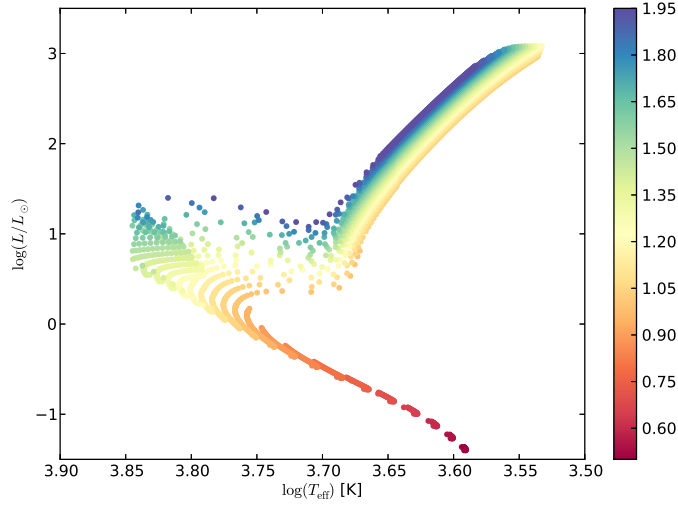}}
\caption{Position of the stellar modes in the Hertzsprung-Russell diagram. The color of the symbols correspond to masses in solar units of the models, and the corresponding color scale is given in the vertical color bar.}
\label{fig:HR}
\end{figure}

As depicted in Fig.~\ref{schema_deltanu}, to infer the sources of uncertainties of the $\Delta \nu_{\rm obs} - \nu_0$, it is useful to consider separately the $\Delta \nu_{\rm obs} - \Delta \nu_{\rm as}$ and $\Delta \nu_{\rm as} - \nu_0$ relations. We therefore quantified, with the help of a grid of stellar models, the departure from a 1:1 scaling of both relations.  

\subsubsection{Grid of stellar models}
\label{grid}

The grid was computed with  the CESTAM code \citep{Marques2013} and includes stellar models with masses ranging from $M=0.5\,M_\odot$ to $M= 1.95\,M_\odot$ (see Fig.~\ref{fig:HR}). 
The models include standard physics and  do not include microscopic diffusion nor rotation. They  all have a solar metal abundance assuming \citet{Asplund05} chemical mixture (i.e. $Y=0.2485$ and  $Z/X=0.165$). The positions of the models in the Hertzsprung-Russell diagram are shown Fig.~\ref{fig:HR}. The  theoretical mode frequencies associated with the stellar models are obtained with the ADIPLS code \citep{JCD2011_adipls}. The stellar models were re-meshed with an adapted grid of 8\,000 points. We consider only radial order modes.  For each stellar model, radial eigenfrequencies are used to compute $\dnuobs$. For $\dnuobs$ we proceed in practice in a similar way as \citet{White2011}, \emph{i.e.} we determine $\dnuobs$ by adjusting by means of least-square the first-order asymptotic relation 
\eqn{
\nu_{n,0} = \dnuobs \, \left ( n + \epsilon \right ) \, ,
} 
where $\nu_{n,0}$ is the radial mode eigenfrequency, $n$ the associated radial order, and $\epsilon$ an offset. The weight entering the least square fit is a Gaussian function centered on $\numax$. The FWHM of the Gaussian function, $\delta \nu_{\rm env}$ is assumed to depend on $\numax$. For stars with $\numax < 200~\mu$Hz (RG stars), we adopt the scaling relation obtained by \citet{Mosser2012a}, while above $\numax = 200~\mu$Hz,  $\delta \nu_{\rm env}$ is assumed to scale linearly with $\numax$. 

\begin{figure}
\center
\includegraphics[width=9cm]{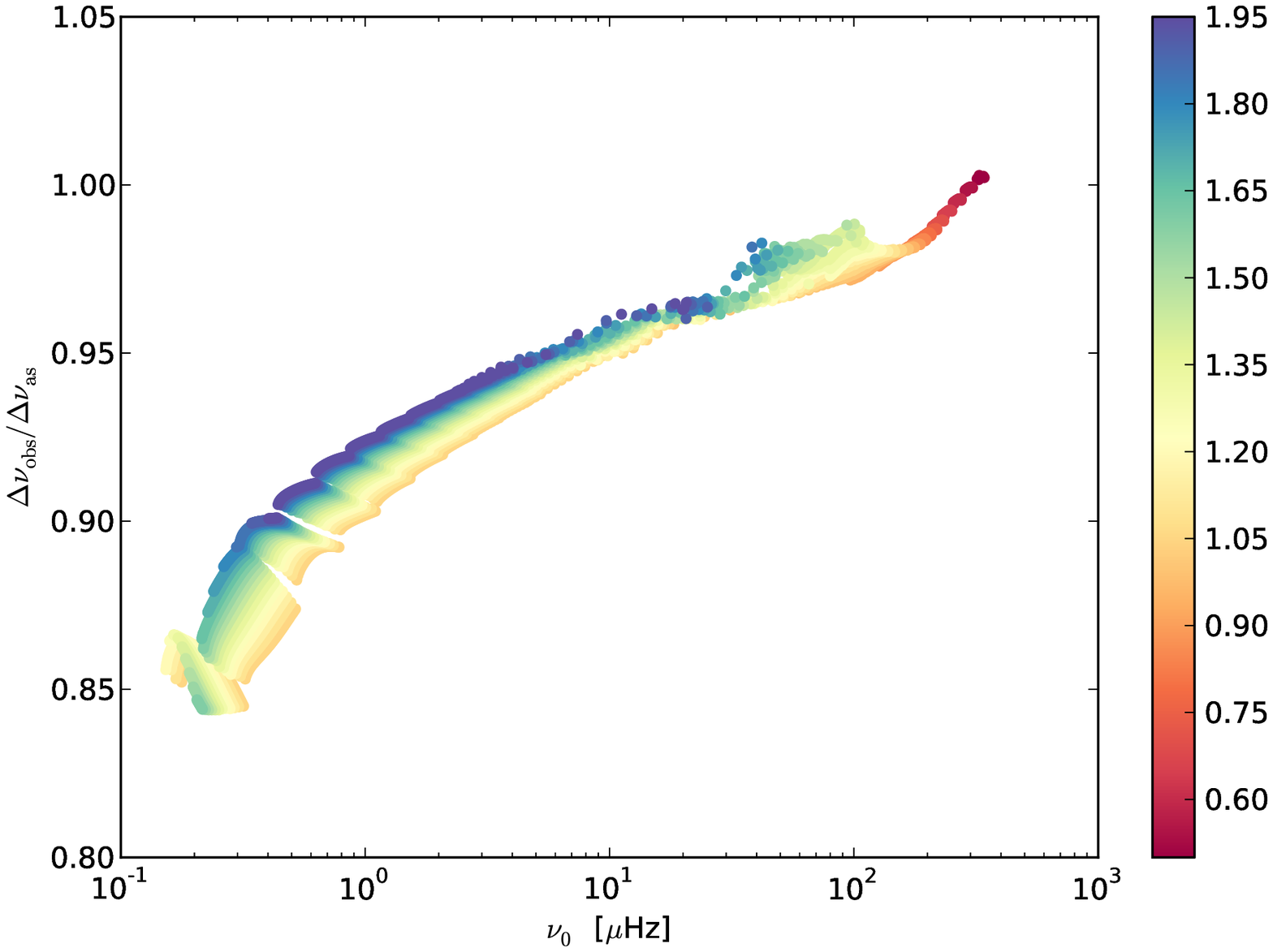}
\includegraphics[width=9cm]{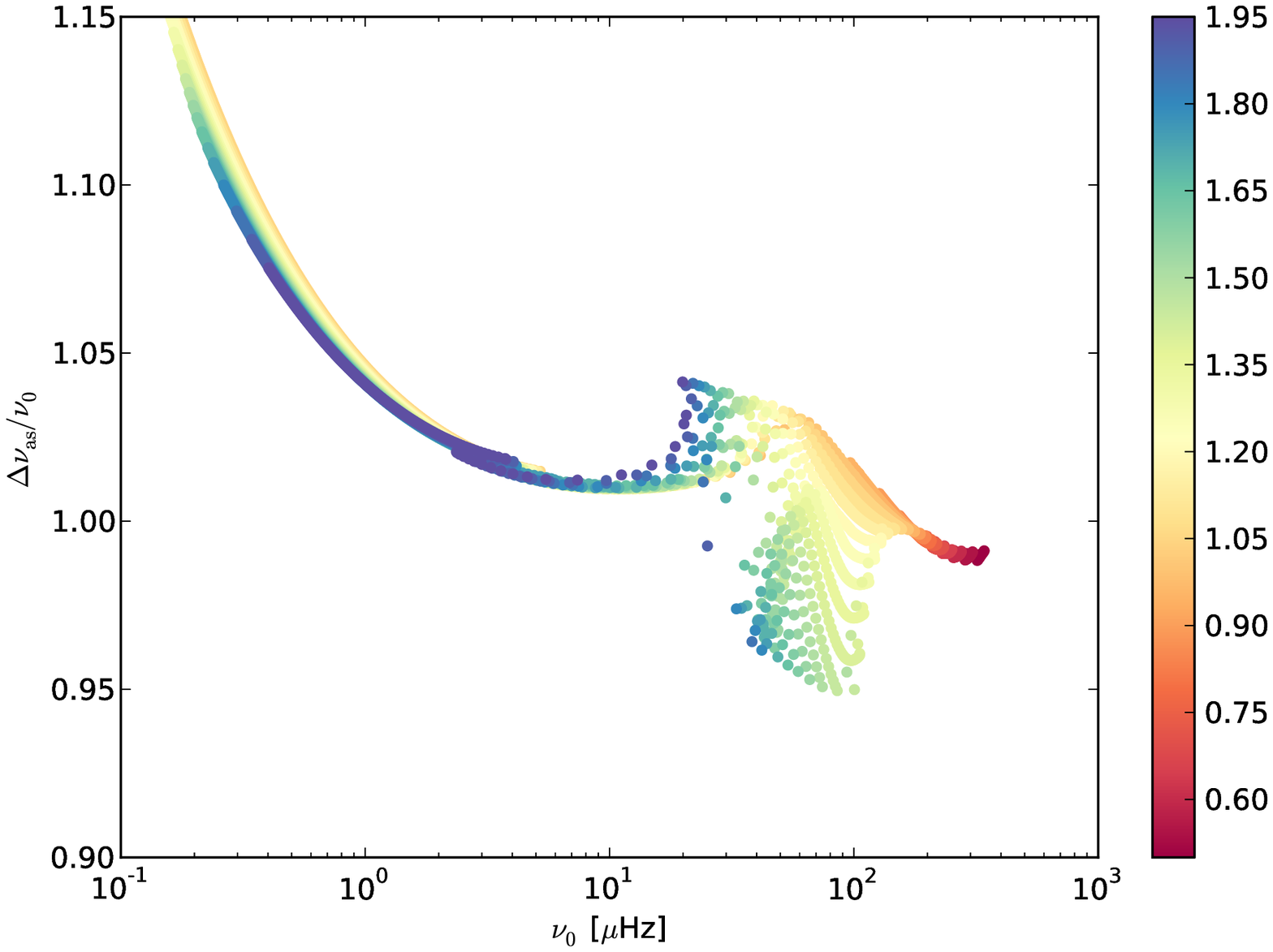}
\caption{{\bf Top:} $\left(\dnuobs / \dnuas \right)$ as a function of $\nu_0$ for stellar models as described in Sect.~\ref{grid} and summarized in the HR diagram (Fig.~\ref{fig:HR}). Same color code as in Fig.~\ref{fig:HR}. {\bf Bottom:} The same as for the top panel but for the ratio $\left( \dnuas / \nu_0 \right)$.}
\label{fig:departure_sources}
\end{figure}

\subsubsection{The $\Delta \nu_{\rm obs} - \Delta \nu_{\rm as}$ relation}

Let us first consider the intermediate relation between $\dnuobs$ and $\dnuas$. As shown by Fig.~\ref{fig:departure_sources} (top panel), the departure from unity of the ratio $\dnuobs / \dnuas$ is of several percent for main-sequence stars and early sub-giant and can reach up to 15 \% for red giant stars near the tip of the branch. We also note that the dispersion remains small. Therefore, this departure is only weakly mass dependent. 

There are several physical reasons that can explain such a departure from a perfect scaling, namely: the departure from the asymptotic expansion (\emph{i.e.} the conditions $n \gg 1$ is not fulfilled), glitches (\emph{i.e.} the effect of discontinuities of the sound speed profile), or surface effects (\emph{e.g.} effects of turbulent pressure and non-adiabatic processes that are not considered in the computations).  

Among all those possible biases, the departure from the asymptotic regime seems the dominant one for red giant stars. Indeed, while $n_{\rm max} \simeq 20$ for main-sequence stars, it decreases as the stars evolve on the subgiant and red giant phases to reach up to $n_{\rm max} \simeq 2$ near the tip of the branch. A correction of this effects has recently been proposed by \cite{Mosser2013} \citep[see also][]{Hekker2013}. It provides an estimate of the bias between  $\Delta \nu_{\rm obs}$ and  $\Delta \nu_{\rm as}$, which is in qualitative agreement with Fig.~\ref{fig:departure_sources} (top panel). 

More insights are nevertheless desirable to fully understand the departures between $\Delta \nu_{\rm obs}$ and  $\Delta \nu_{\rm as}$. This will need to investigate the effects of varying the input physics of the models and this should help to fine-tune prescriptions \citep[such as proposed by][]{Mosser2013} to relate $\Delta \nu_{\rm obs}$ and $\Delta \nu_{\rm as}$. 

\subsubsection{The $\Delta \nu_{\rm as} -  \nu_{0}$ relation}

The ratio $\Delta \nu_{\rm as} / \nu_{0}$ as a function of $\nu_0$ is displayed in Fig.~\ref{fig:departure_sources} (bottom panel). One immediately sees that the departure from a perfect scaling is up to 5 \% for main-sequence stars, with an important dispersion in mass, and can reach up to 10 \% for red-giant stars, but with a small dispersion in mass. 

From a physical point of view, these behaviors can be understood on the basis of the homology as already presented in Sect.~\ref{homology}. For main-sequence stars, $\Delta \nu_{\rm as}$ remains sensitive to the physical conditions in the core. This explains the dispersion in mass that reflects the variations of the physical conditions along the main-sequence phase with the mass. For instance, as the mass increases, the nuclear reactions switch from a p-p chain to a CNO chain (inducing a change in the mechanism of transport of energy from a radiative to a convective transport in the core). This of course violates the conditions for homology to apply, as discussed in Sect.~\ref{homology}. 

For sub-giant stars and giant stars the situation is quite different. Indeed, as the star evolves, its core contracts and its envelope dilates, so that the integral involved in the computation of $\Delta \nu_{\rm as}$ mainly depends on the convective upper layers. Adiabatic convection is reasonably well modeled by a polytrope of index $1.5$ and it is well known that two polytropes of same index are homologous \cite[e.g., ][]{Chandrasekhar67}. Consequently, one can infer that the departure of a ratio $\Delta \nu_{\rm as} /  \nu_{0}$ from unity and its trend are dominated by the increasing influence of the super-adiabatic layers as a star evolves. 
 
Once again, a more dedicated work on this issue will help to improve the accuracy of the $\Delta \nu_{\rm as} -  \nu_{0}$ relation.  

\subsection{Discussion}

\begin{figure}
\centerline{\includegraphics[width=10cm]{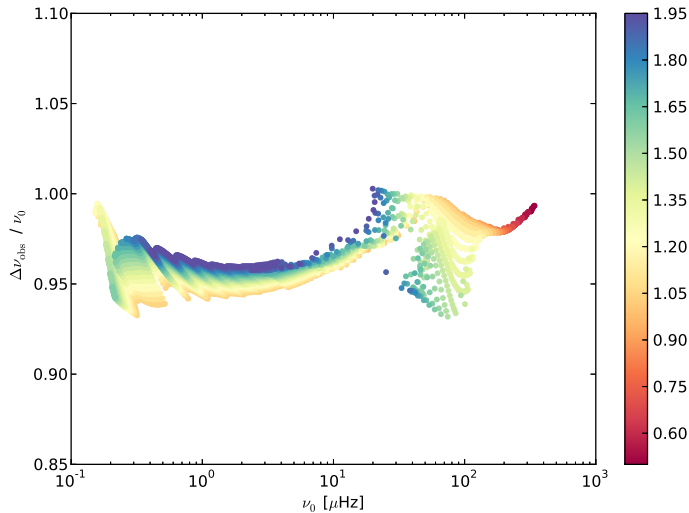}}
\caption{ Ratio $\dnuobs/\nu_0$ as a function of $\nu_0$ (see Sect.~\ref{definitions_prealables} for the definitions). The grid of models is the same as in Fig.~\ref{fig:HR}. Same color code as in Fig.~\ref{fig:HR}.}
\label{fig:delta_obs}
\end{figure}

Finally, the departure of $\dnuobs$ from $\nu_0$ shown in Fig.~\ref{fig:delta_obs} is up to 5\,\%, with a strong dependence on mass for main-sequence stars and sub-giants stars, while for RGB stars with $\nu_0 \lesssim 600\,\mu$Hz, this departure depends more weakly on the mass  and --~ in average ~--  slowly increases with decreasing $\nu_0$.  It is striking to note that there is a smaller departure of  $\dnuobs$  from $\nu_0$, which is mainly explained by the compensation effect of the departure of the mode frequencies from the asymptotic regime (\emph{i.e.} from the $\dnuobs - \dnuas$ relation) with the departure from the homology (\emph{i.e.} from the $\dnuas - \nu_0$ relation). 

Whether this is a coincidence or can be explained by some fundamental reasons remains an open issue. Anyway, it is quite a good news since it justifies the extensive use of the $\Delta \nu_{\rm obs} - \nu_0$ scaling relation. Moreover, it allows the use of $\Delta  \nu_{\rm obs}$ to derive stellar parameters and more particularly for red-giant stars, for obtaining a proxi of the stellar mass and radius. 

\section{The $\numax-\nu_{\rm c}$ scaling relation}
\label{sect:numax}

\begin{figure}
\centerline{\includegraphics[height=6.cm,width=8cm]{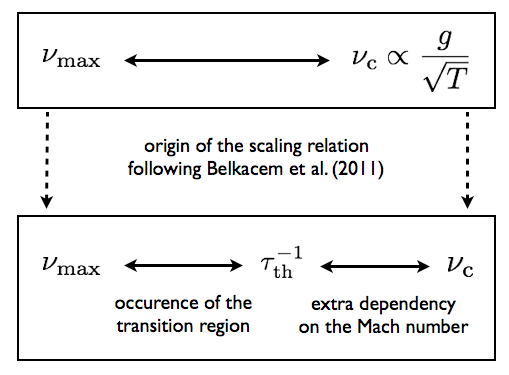}}
\caption{Sketch that illustrates the relation between $\numax$ and $\nuc$ (top panel). Following the work of \cite{Belkacem2011}, the bottom panel explicits the intermediate frequency that is the thermal frequency. }
\label{schema_numax}
\end{figure}

As introduced in Sect.~\ref{intro}, the scaling relation that provides an estimate of the surface gravity derives from the proportionality between $\nu_{\rm max}$ and $\nu_{\rm c}$. In this section, we discuss the theoretical foundations of this scaling relation.  We follow the work of \cite{Belkacem2011}, which shows that this relation can be explained by two intermediate relations, namely $\numax - \tau_{\rm th}^{-1}$ (where $ \tau_{\rm th}^{-1}$ is the thermal frequency, see Sect.~\ref{relation_nu_th} for a precise definition) and the $\tau_{\rm th}^{-1}-\nuc$, as displayed in Fig.~\ref{schema_numax}. 

\subsection{The transition region and the $\numax-\nu_{\rm th}$ relation}
\label{sect:transition}

The frequency $\numax$ is determined by the maximum height $H$ of the background-corrected power spectrum. For stochastically excited modes, the height of the mode profile in the power spectrum is given by \citep[e.g.][]{Libbrecht1988,Chaplin1998,Baudin2005,Chaplin2005,Belkacem2006}
\begin{equation}
\label{heighteqres}
H\;=\;\frac{P\, }{2\,\eta^2\,\mathcal{M}} \, , \quad {\rm with} \quad \mathcal{M} = \int_{0}^{M} \frac{\vert \xi \vert^2}{ \vert \xi (M) \vert^2} {\rm d}m \, , 
\end{equation}
where $P$ is the excitation rate, $\eta$ is the damping rate, $\mathcal{M}$ is the mode mass, and $\xi$ is the mode displacement. As shown for instance by \cite{Chaplin2008,Belkacem2011}, and confirmed with  observations of the solar-like stars by \emph{Kepler} \citep{Appourchaux2012}, the maximum of $H$ is predominantly determined by the squared damping rates $(\eta^2)$ in Eq.~(\ref{heighteqres}). More precisely, $\numax$ arises from the depression (or plateau) of $\eta$. 

\begin{figure}
\centerline{\includegraphics[height=6cm,width=9cm]{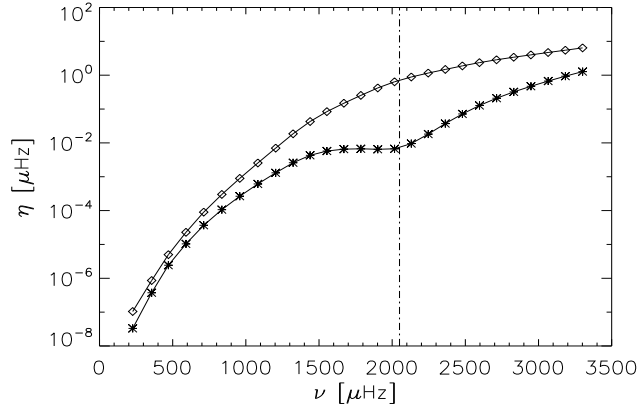}}
\caption{Mode damping rates versus mode frequency computed for a model of one solar mass on the main-sequence, using the \cite{MAD05} formalism as described in \cite{Belkacem2012}. The star symbols correspond to the full computation while the diamond symbols correspond to the computation for which we imposed $\delta \kappa / \kappa = 0$. 
The vertical dashed-dotted line corresponds to the frequency $\numax$ computed using the scaling relation (Eq.~\ref{scaling_numax}).}
\label{fig_transition}
\end{figure}

The depression of the damping rates occurs when the modal period nearly equals the thermal time-scale (or thermal adjustment time-scale) in the superadiabatic layers.  This was first mentioned by \cite{Balmforth92} (see his Sect. 7.2 and 7.3) and confirmed by \cite{Belkacem2011}, on the basis of two different non-adiabatic pulsation codes. In the context of classical pulsators, the location of this equality is referred to as the transition region \citep[e.g.][]{Cox74,Cox80} and its occurrence in the ionization region is one of the necessary condition for a mode to be excited by the $\kappa$-mechanism \citep[e.g.,][]{Cox80,CoxGuili68,Pam99}. In the context of solar-like pulsators, the situation is very similar, except that the destabilization by the perturbation of the opacity never dominates over damping terms \citep{Belkacem2012} and that the situation is complicated by the presence of convection which modifies the thermal time-scale \citep[see][for details]{Belkacem2011,Belkacem2012b}. This is illustrated in Fig.~\ref{fig_transition}, which  displays the mode damping rates computed using the \cite{MAD05} formalism. It confirms that the perturbation of the opacity is the corner-stone of the relation between the modal period and the thermal time-scale. 

\subsection{The $\nu_{\rm th}-\nuc$ relation from 3D numerical simulations}
\label{relation_nu_th}

As shown in the previous section, there is a linear relation between $\numax$ and the thermal frequency $\nu_{\rm th} \equiv \tau_{\rm th}^{-1}$. Let us now investigate the relation between $\nu_{\rm th}$ and $\nu_{\rm c}$. 

The thermal adjustment time-scale has been extensively discussed by \cite{CoxGuili68,Cox80,Pesnell83}. It is defined as 
\begin{eqnarray}
\label{thermal_time}
\tau_{\rm th} = \frac{1}{L} \int_{m_{\rm tr}}^{M} c_v T {\rm d}m
\end{eqnarray}
where $M$ is the total mass, $c_v$ is specific heat capacity at fixed volume, and $m_{\rm tr}$ is the mass at the transition region. 
As already explained in Sect.~\ref{sect:transition}, the relation between $\numax$ and $\nu_{\rm th}$ holds due to the occurrence of the transition region in the ionization region. Accordingly, we compute lower boundary of the integral in \eq{thermal_time} as the minimum of $(\Gamma_3-1)= (\partial \ln T / \partial \ln \rho )_s $, where $s$ is the specific entropy\footnote{Note that the scaling is hardly sensitive to this lower boundary.}. 

\begin{figure}
\centerline{\includegraphics[height=7cm,width=9cm]{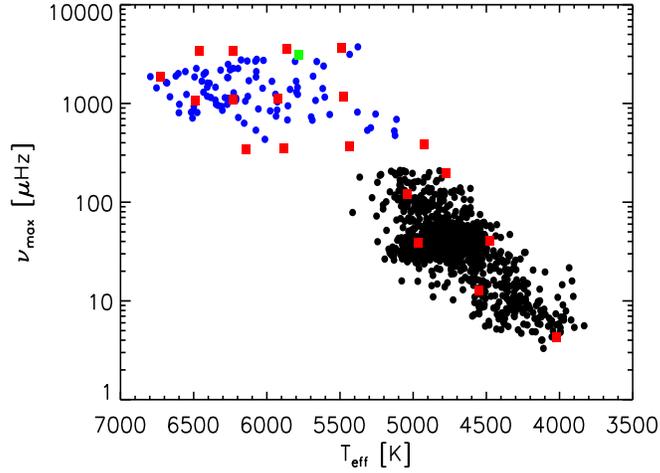}}
\caption{$\numax$ as a function of the effective temperature ($T_{\rm eff}$). The filled red squares corresponds to the location of the 3D hydrodynamical models (see text for details), the filled blue circles to sub-giant and main-sequence targets observed by \emph{Kepler} \citep{Chaplin2011}, and the black ones to the red-giant stars observed by \emph{Kepler} \citep{Mathur2011}. Finally, the green square corresponds to the solar 3D model. }
\label{fig_simu3D_HR}
\end{figure}

For the cut-off frequency, a general expression has been proposed by \cite{Balmforth90} 
\begin{equation}
\omega_c = 2\pi \nuc = \left( \frac{c_s}{2 H_\rho} \right) \sqrt{1-2 \deriv{H_\rho}{r}}
\end{equation}
with $c_s$ the sound speed, and $H_\rho = -({\rm d}\ln \rho / {\rm d}r)^{-1}$ the density scale height. For an isothermal atmosphere, $\omega_c$ reduces to 
\begin{equation}
\omega_c = \frac{c_s}{2 H_\rho} \, .
\end{equation}
To go further it is customary to use the pressure scale height ($H_p$) as a proxy for the density scale height. This is based on the fact that both quantities are nearly equal at the photosphere. Finally, it can be shown that $H_p$ scales as the ratio between the surface gravity and the square root of the effective temperature. 
Consequently, in the following, we will denote and compute the cut-off frequency as 
\begin{equation}
\omega_c = \frac{c_s}{2 H_p} \propto \frac{g}{\sqrt{T_{\rm eff}}} \, , 
\end{equation}
where we considered all the quantities at the photosphere, and we made use of the scaling relations $c_s^2 \propto T_{\rm eff}$, $H_p \propto T_{\rm eff} / g$. 

\begin{figure}
\centerline{\includegraphics[height=7cm,width=9cm]{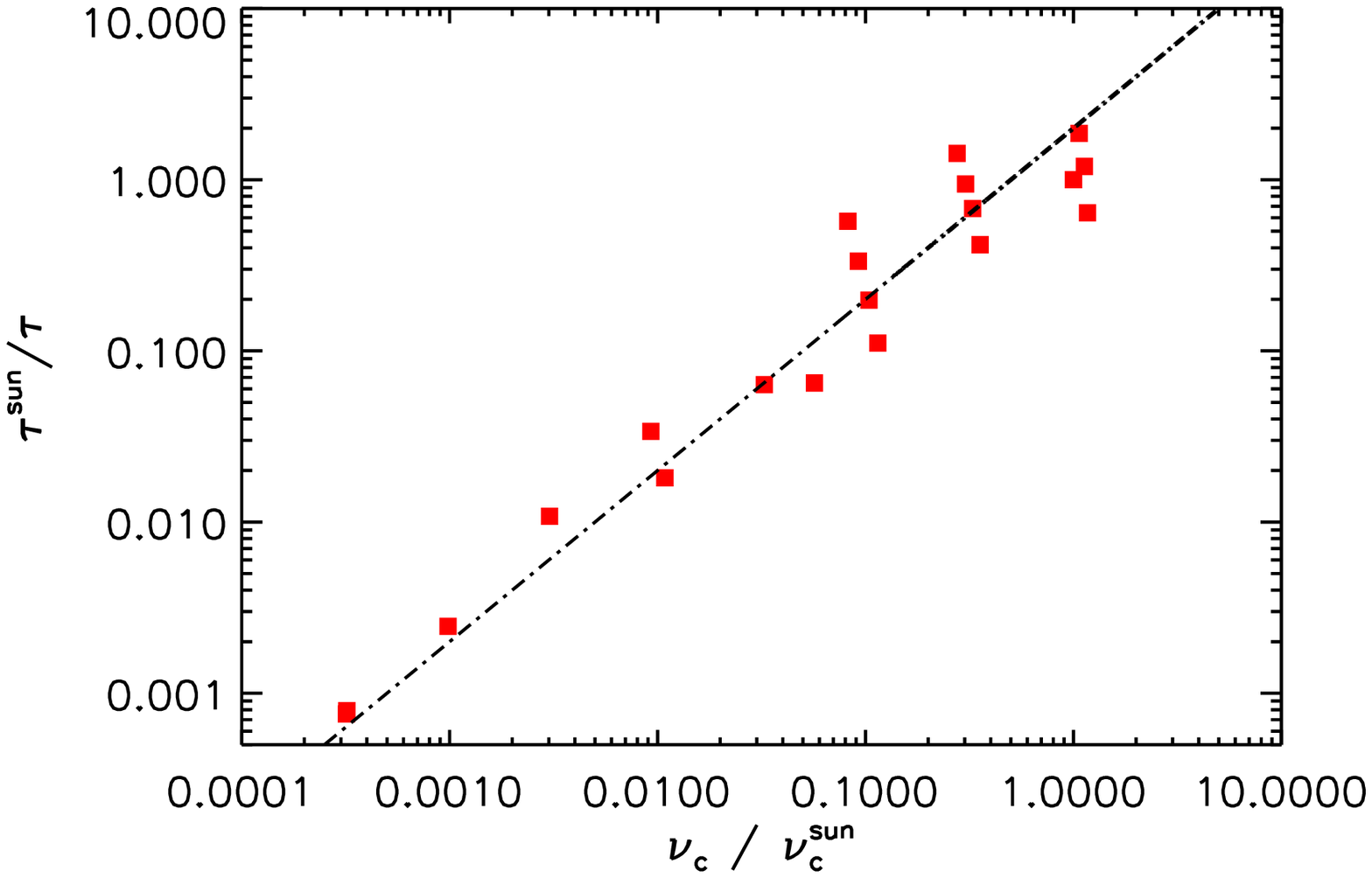}}
\centerline{\includegraphics[height=7cm,width=9cm]{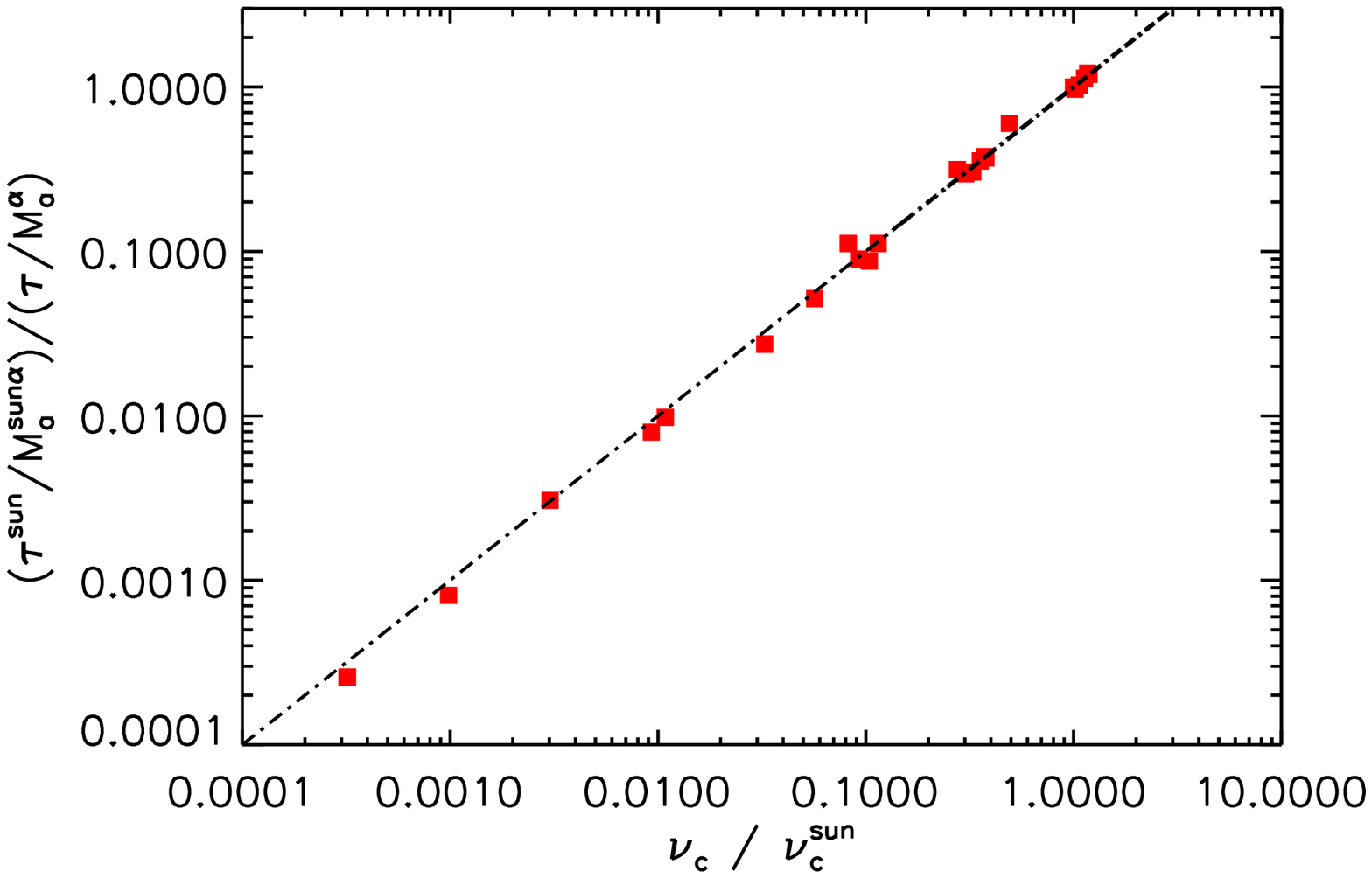}}
\caption{\emph{Top panel:}  Thermal frequency $\tau_{\rm th}^{-1}$ as a function of the cut-off frequency $\nuc$. All the quantities are normalized by the values derived from the Solar 3D simulation. The filled squared correspond to the 3D models displayed in Fig.~\ref{fig_simu3D_HR}. The dashed-dotted line corresponds to the linear curve. \emph{Bottom panel:} As for the top panel, except that the thermal frequency is corrected by the term $\mathcal{M}_a^\alpha$ with $\alpha = 2.78$ (see \eq{numax_mach}).}
\label{fig_simu3D_results}
\end{figure}

\cite{Belkacem2011} used the mixing-length formalism to derive a linear relation between $\nu_{\rm th}$ and $\nu_{\rm c}$. However, it is well-known that the MLT suffers many deficiencies and particularly near the photosphere. Therefore, we propose here to employ a set of 3D hydrodynamical numerical simulations. To this end, we used 3D models from the CIFIST grid \citep{Ludwig09b}. The CIFIST grid covers the main sequence and the giant branch of late-type stars. All  3D models used here have a solar metal abundance with a chemical mixture similar to the solar chemical composition proposed by \citet{Asplund05}. Their characteristics are extensively described in \cite{Samadi2013b}.  We determine for each 3D model the associated radius using a grid of standard stellar models computed with the CESTAM code \citep{Marques2013}. The stellar 1D models have the same chemical composition than the 3D models.

Figure~\ref{fig_simu3D_HR} shows how the selected simulations span into the $\log g - T_{\rm eff}$ (which translates in the $\nu_{\rm max} - T_{\rm eff}$ from an observational point of view) diagram and how it compares with the recent {\it Kepler} and CoRoT observations. Our 3D models are representative of the current observations of solar-like pulsators from the main-sequence to the red-giant phase. 
The linear relation derived by \cite{Belkacem2011} is confirmed by the 3D models, even if there is some dispersion especially for main-sequence stars (see the top panel of Fig.~\ref{fig_simu3D_results}). In addition, we confirm that the main source of uncertainty is related to an extra factor that is the Mach number ($\mathcal{M}_a$). More precisely, let us assume that
\begin{equation}
\label{numax_mach}
\tau_{\rm th}^{-1} \propto \mathcal{M}_a^\alpha \; \nu_{\rm c} \, , 
\end{equation}
so that one can derive $\alpha$ to minimize the dispersion. It gives
\begin{equation}
\alpha = 2.78 
\end{equation}
This result agrees with what can be derived from the mixing length theory, \emph{i.e.} $\alpha=3$. Indeed, this is quite a robust number since the dependence to the Mach number can be derived from simple energetical arguments that hardly depend on the assumptions related to the MLT. The resulting scaling relation, that accounts for the dependence on the Mach number, is depicted in Fig.~\ref{fig_simu3D_results} (bottom panel). It confirms that most of the dispersion of the $\nu_{\rm th}-\nuc$ relation (and therefore the $\numax-\nuc$ relation) comes from the extra dependence on $\mathcal{M}_a$. Note that further investigations on the influence of the P\'eclet number (ratio of the radiative to the convective time-scales)  would be desirable \citep[see][]{Tremblay2013}.

\subsection{Effect of the Mach number on the scaling for stars on the red giant branch}

In this section, our objective is to investigate why the effect of the Mach number seems to be negligible on the $\numax-\nuc$ relation for red-giant stars, as shown in Fig.~\ref{fig_simu3D_results} (top panel). To this end, we will first make several assumptions that will enable us to derive a scaling between the Mach number and stellar global parameters.  First we assume that the total flux is convective and proportional to the kinetic energy flux,  so that 
\begin{equation}
\label{scaling_mach_tmp}
\mathcal{M}_a \propto T_{\rm eff}^{5/6} \, \rho^{-1/3} \, , 
\end{equation}
where $\rho$ is the surface density. To go further, we note that the surface density does not scale as the mean density $\bar{\rho}$. We therefore use the fact that the optical depth is roughly $\tau \simeq \kappa \rho H_p \equiv 2/3$ and that the opacity is dominated by H$^-$ so that $\kappa \propto \rho^{1/2}\, T^9$ \cite[e.g.][]{HansenKawaler1994}. Thus, \eq{scaling_mach_tmp} becomes
\begin{equation}
\label{scaling_mach}
\mathcal{M}_a \propto T_{\rm eff}^{3} \, g^{-2/9} \, , 
\end{equation}
where $g$ is the surface gravity. 

\begin{figure}
\centerline{\includegraphics[height=7cm,width=9cm]{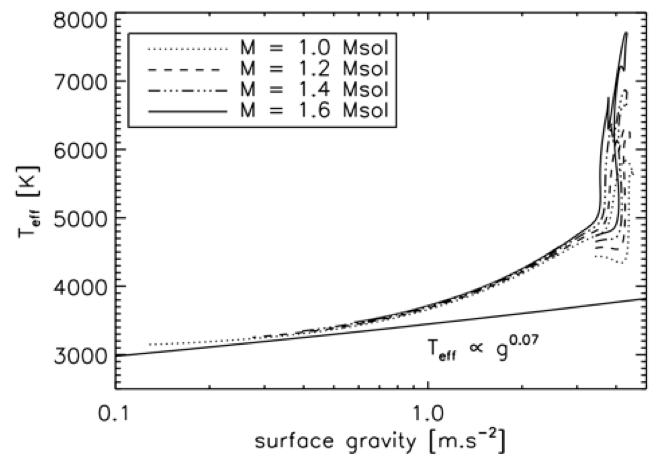}}
\caption{Diagram of the effective temperature versus surface gravity for models computed with the CESTAM code from the main-sequence to the red-giant branch.} 
\label{fig_grid_teff_g}
\end{figure}

As shown in Fig.~\ref{fig_grid_teff_g}, there is an additional relation between the surface gravity and the effective temperature of stars on the red giant branch. It reads
\begin{equation}
\label{Teff-g}
T_{\rm eff} \propto g^{0.07} \, .
\end{equation}
This is in good agreement with the observations that give $T_{\rm eff} \propto \numax^{0.068}$ \citep{Mosser2013b}. 

It is then possible to understand why the $\numax-\nuc$ relation hardly depends on the Mach number for red giant stars on the red giant branch. Indeed, the Mach number becomes nearly  independent of both effective temperature and surface gravity. If we introduce \eq{scaling_mach} and \eq{Teff-g} into \eq{numax_mach}, we obtain
\begin{equation}
\nu_{\rm max} \propto \mathcal{M}_a^3 \, \nu_{\rm c} \propto  \frac{g^{0.988}}{\sqrt{T_{\rm eff}}} \approx {\rm cste} \;\times  \nu_{\rm c} \, .
\end{equation}
This result shows that for red-giant stars near the tip of the branch, the effect of the Mach number becomes negligible. In other words, one can conclude that the $\numax - \nuc$ relation is more accurate for red giants since the possible biases introduced by the Mach number become small. 

\section{Concluding remarks}
\label{sect:conclu}

In this paper, we discussed the physical meaning of the scaling relations $\Delta \nu - \bar{\rho}$ and $\numax - \nuc$, the foundations of what is now commonly called \emph{Ensemble Asteroseismology}. 

We have discussed the $\Delta \nu - \bar{\rho}$ relation with emphasis on the possible departure from a linear relation. It turns out that the relation holds within several percents from the main-sequence to the red giant phase. This validity is obtained thanks to compensating effects mainly between departure from the homology and from the asymptotic regime. Concerning the $\numax - \nuc$ relation, we confirm the physical interpretation of \cite{Belkacem2011} by using a set of of 3D hydrodynamic models representative for the CoRoT and {\it Kepler} observations.  

Finally, one must note that, at this stage of our physical understanding of the $\Delta \nu - \bar{\rho}$ and $\numax - \nuc$ scaling relations, it is not yet possible to firmly conclude about their accuracy. 

\acknowledgements K. B.  is  grateful to the organizers for their invitation and for providing financial support. HGL acknowledges financial support by the Sonderforschungsbereich SFB\,881 "The Milky Way System'' (subproject A4) of the German Research Foundation (DFG). We also acknowledge financial support from "Programme National de Physique Stellaire" (PNPS) of CNRS/INSU, France. The authors also acknowledge financial support from the French National Research Agency (ANR) for the project ANR IDEE (Influence Des Etoiles sur les Exoplan\`etes). 

\bibliography{belkacem}
\end{document}